\documentclass[10pt,conference]{IEEEtran}
\IEEEoverridecommandlockouts
\usepackage{cite}
\usepackage[numbers]{natbib}
\usepackage{amsmath,amssymb,amsfonts}
\usepackage{algorithmic}
\usepackage{graphicx}
\graphicspath{{./figs/}}
\DeclareGraphicsExtensions{.pdf,.jpeg,.png,.eps}
\usepackage{textcomp}
\usepackage{xcolor}
\usepackage{caption}
\usepackage{subcaption}
\usepackage[colorlinks,bookmarksopen,bookmarksnumbered,citecolor=red, urlcolor=black]{hyperref}
\usepackage{booktabs, caption, tabularx}
\usepackage{fancybox}  % for \hypobox
\usepackage{multirow}
\usepackage{xltabular}
\usepackage{xspace}
\usepackage{xcolor}

\DeclareTextFontCommand{\emp}{\bfseries}

\newcommand{\hypobox}[1]{\begin{center}%
		\noindent\thicklines\setlength{\fboxsep}{8pt}%
		\cornersize{0.0}\Ovalbox{\begin{minipage}{3in}%
				#1\end{minipage}} \end{center}}

\newcommand{\motivation}{\emp{Motivation. }}
\newcommand{\approach}{\emp{Approach. }}
\newcommand{\results}{\emp{Results. }}

\def\BibTeX{{\rm B\kern-.05em{\sc i\kern-.025em b}\kern-.08em
    T\kern-.1667em\lower.7ex\hbox{E}\kern-.125emX}}
\begin{document}
\bstctlcite{IEEEexample:BSTcontrol} % make this the first directive after "\begin{document}"

% Candidates:
% Achieving Robustness in Deep Learning Models through History-Based Overfitting Mitigation
% Enhancing Deep Learning Model Quality: A History-Based Approach to Overfitting Detection and Prevention
% Improving Model Trustworthiness: A History-Based Approach to Overfitting Detection and Prevention
% Enhancing Model Trustworthiness: A History-Based Approach to Detect and Prevent Overfitting in Deep Learning
% Enhancing Deep Learning Model Reliability: A History-Based Overfitting Guard
% Overfitting awareness: An approach to enhance the quality of deep learning models based on training history curves.
% Keeping Deep Learning Models on Track: A History-Based Overfitting Detection and Prevention Approach

% \title{OverfitGuard: A History-Based Model Validation Approach to Detect and Prevent Overfitting}
\title{Keeping Deep Learning Models in Check: A History-Based Approach to Mitigate Overfitting}
% \title{A History-Based Model Validation Approach to Detect and Prevent Overfitting in Software Systems}

% \author{\IEEEauthorblockN{Hao Li}
% \IEEEauthorblockA{\textit{University of Alberta}\\
% li.hao@ualberta.ca}
% \and
% \IEEEauthorblockN{Gopi Krishnan Rajbahadur}
% \IEEEauthorblockA{\textit{Huawei Canada}\\
% gopi.krishnan.rajbahadur1@huawei.com}
% \and
% \IEEEauthorblockN{Dayi Lin}
% \IEEEauthorblockA{\textit{Huawei Canada}\\
% dayi.lin@huawei.com}
% \and
% \IEEEauthorblockN{Cor-Paul Bezemer}
% \IEEEauthorblockA{\textit{University of Alberta} \\
% bezemer@ualberta.ca}
% \and
% \IEEEauthorblockN{Zhen Ming (Jack) Jiang}
% \IEEEauthorblockA{\textit{York University} \\
% zmjiang@cse.yorku.ca}
% }

\author{\IEEEauthorblockN{Hao Li\IEEEauthorrefmark{1},
Gopi Krishnan Rajbahadur\IEEEauthorrefmark{2}, Dayi Lin\IEEEauthorrefmark{2},
Cor-Paul Bezemer\IEEEauthorrefmark{1}, and Zhen Ming (Jack) Jiang\IEEEauthorrefmark{3}}
\IEEEauthorblockA{
\IEEEauthorrefmark{1}University of Alberta. \{li.hao, bezemer\}@ualberta.ca,\\
\IEEEauthorrefmark{2}Centre for Software Excellence, Huawei Canada. \{gopi.krishnan.rajbahadur1, dayi.lin\}@huawei.com,\\
\IEEEauthorrefmark{3}York University. zmjiang@cse.yorku.ca}}
% \author{Anonymous Authors}

\maketitle

\begin{abstract}
In software engineering, deep learning models are increasingly deployed for critical tasks such as bug detection and code review. However, overfitting remains a challenge that affects the quality, reliability, and trustworthiness of software systems that utilize deep learning models. Overfitting can be (1)~prevented (e.g., using dropout or early stopping) or (2)~detected in a trained model (e.g., using correlation-based approaches). Both overfitting detection and prevention approaches that are currently used have constraints (e.g., requiring modification of the model structure, and high computing resources). In this paper, we propose a simple, yet powerful approach that can both detect and prevent overfitting based on the training history (i.e., validation losses). Our approach first trains a time series classifier on training histories of overfit models. This classifier is then used to detect if a trained model is overfit. In addition, our trained classifier can be used to prevent overfitting by identifying the optimal point to stop a model's training. We evaluate our approach on its ability to identify and prevent overfitting in real-world samples. We compare our approach against correlation-based detection approaches and the most commonly used prevention approach (i.e., early stopping). Our approach achieves an F1 score of 0.91 which is at least 5\% higher than the current best-performing non-intrusive overfitting detection approach. Furthermore, our approach can stop training to avoid overfitting at least 32\% of the times earlier than early stopping and has the same or a better rate of returning the best model. %We also provide guidelines for software engineers on how to use our approach seamlessly as part of their software system (which trains or uses deep learning models) with minimal performance trade-offs.

% Overfitting in deep learning models negatively impacts their generalizability on unseen data, consequently affecting the quality of software systems that utilize these trained models. Overfitting can be (1)~prevented (e.g., using dropout or early stopping) or (2)~detected in a trained model (e.g., using correlation-based approaches). We propose an approach that can both detect and prevent overfitting based on the training history (i.e., validation losses). Our approach first trains a time series classifier on training histories of overfit models. This classifier is then used to detect if a trained model is overfit. In addition, our trained classifier can be used to prevent overfitting by identifying the optimal point to stop a model's training. We evaluate our approach on its ability to identify and prevent overfitting in real-world samples (collected from papers published in the last 5 years at top AI venues). We compare our approach against correlation-based detection approaches and the most commonly used prevention approach (i.e., early stopping). Our approach achieves an F1 score of 0.91 which is at least 5\% higher than the current best-performing non-intrusive overfitting detection approach. Furthermore, our approach can stop training to avoid overfitting at least 32\% of the times earlier than early stopping and has the same or a better rate of returning the best model.

\end{abstract}


\newcommand{\OverfitGuard}{\texttt{OverfitGuard}\xspace}
\newcommand{\rqone}{How well does \OverfitGuard detect overfitting in trained DL models?}
\newcommand{\rqtwo}{How well does \OverfitGuard prevent overfitting during the training process?}

\section{Introduction}

% Overfitting is one of the fundamental issues that plagues the field of machine learning~\citep{nowlan_first_1992, ng_overfit_1997, caruana2000overfitting, cawley2007preventing, erhan_why_2010, srivastava2014dropout, zhao_multiple_2020}. Overfitting can also occur when training a deep learning (DL) model and has negative implications for the quality of machine learning software systems. In essence, overfitting is defined as unknowingly extracting some of the residual variation (i.e., noise) as if the variation represented the underlying model structure~\citep{anderson_overfit_2004, hawkins2004problem}. An overfit model increases the risk of inaccurate predictions, misleading feature importance, and wasted resources~\citep{hawkins2004problem}. Figure~\ref{fig:examples} shows example training histories~(i.e., the training and validation losses curves) of an overfit and a non-overfit model. The training and validation losses of the overfit model both decrease at the beginning of the training process. Following that, the validation loss increases while the training loss decreases, resulting in a large gap between the training and validation losses. Such a trend indicates that the trained model does not generalize well to new data.

The use of Deep Learning (DL) models in software engineering (SE) research and software products has been skyrocketing over the past decade. For instance, DL techniques have been used for automated bug detection~\citep{hanam_bug_2016}, code review~\citep{sghaier_review_2023}, and software testing~\citep{tian_deeptest_2018}. These applications underscore the importance and ubiquity of DL in modern SE.

Overfitting is one of the fundamental issues that plagues DL models~\citep{Kim_understanding_2021, smith_test_2015, nilizadeh_test_2021, yang_survey_2022,watson_survey_2022}. A DL model can be considered overfitting if the model fits just the training data instead of learning the target hypothesis~\citep{watson_survey_2022}. An overfit model increases the risk of inaccurate predictions, misleading feature importance, and wasted resources~\citep{hawkins2004problem}.

Currently, the problem of overfitting is addressed in SE studies that use DL models by either (1)~preventing it from happening in the first place or (2)~detecting it in a trained DL model~\citep{watson_survey_2022,yang_survey_2022}. Overfitting prevention approaches include early stopping~\citep{morgan1989generalization}, data augmentation~\citep{augmentation_shorten_2019}, regularization~\citep{kukavcka2017regularization}, and modifying the DL model by adding dropout layers~\citep{srivastava2014dropout} or batch normalization~\citep{ioffe2015batch}. However, many of these approaches are intrusive and require modifying the data or the model structure and expertise to execute correctly and even then, they may not work. For instance, adding dropout layers, a popularly used overfitting prevention scheme, when set with a lower threshold or when added to the earlier layers may cause unintentional overfitting~\citep{liu2023dropout}. Furthermore, even the non-intrusive prevention approaches such as early stopping incur trade-offs between model accuracy and training time~\citep{prechelt_early_2012}. For example, late stopping when using the early stopping approach may improve model accuracy, but it will also increase training time. Conversely, stopping too early could result in a sub-optimal model performance.

Overfitting detection approaches like k-fold cross-validation, training the DL model with noisy data points and observing if the added noise impacts the DL model's accuracy~\citep{zhang_pv_2019}, checking if the hypothesis of the trained model and the data are independent~\citep{werpachowski_detecting_2019}  can generally be resource intensive and time consuming. For instance,~\citet{bowen_regu_2016} report that training a DL model to find two semantically linkable questions in StackOverflow takes about 14 hours. If one were to conduct a 5-fold cross-validation to detect if the constructed DL model is overfitting, they would have to invest 70 hours, which might be prohibitive in practice. %Other overfitting detection approaches include retraining the DL model with noisy data points and observing if the added noise impacts the DL model's accuracy~\citep{zhang_pv_2019}. Alternatively, some detection approaches check the hypothesis that the trained model and the data are independent, e.g., \citet{werpachowski_detecting_2019} check the hypothesis by comparing the test error with the estimated test error based on adversarial examples of the test set. Similar to model validation approaches, these approaches often require extra computational resources for activities such as generating adversarial examples, retraining the models, and converting the models.

\begin{figure}[t]
\begin{center}
 	\begin{subfigure}[b]{0.49\columnwidth}
          \centering
 		 \includegraphics[width=\columnwidth]{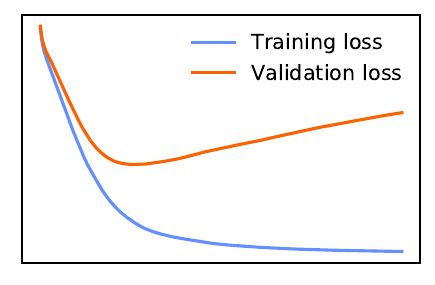}
 		  \caption{Overfitting}
          \label{fig:example_overfit}
 	\end{subfigure}
        % \hfill
 	\begin{subfigure}[b]{0.49\columnwidth}
          \centering
 	 	 \includegraphics[width=\columnwidth]{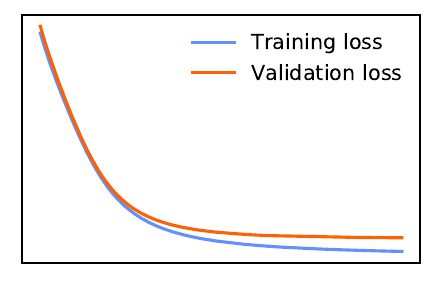}
	 	 \caption{Non-overfitting}
          \label{fig:example_non_overfit}
 	\end{subfigure}
\caption{Examples of overfit and non-overfit training histories.}
\label{fig:examples}
\end{center}
\end{figure}

% In this paper, we introduce \OverfitGuard, a method to detect and prevent overfitting using training histories.
In this paper, we introduce \OverfitGuard, an approach to both detect and prevent overfitting using training histories. Figure~\ref{fig:examples} illustrates example training histories~(i.e., the training and validation losses curves) of an overfit and a non-overfit DL model. The training and validation losses of the overfit model both decrease at the beginning of the training process. Following that, the validation loss increases while the training loss decreases, resulting in a large gap between the training and validation losses. Such a trend indicates poor generalization of the trained model to new data. Researchers have previously employed training histories for decision-making in areas such as quantitative data acquisition and model selection~\citep{rijn2015fast, strang2018curve, bornschein_icml_2020, hoiem_learncurve_2021, mohr2021towards, mohr_learningcurve_2022, brazdil2022metalearning}. Similarly, our approach trains a time series classifier on a simulated dataset of training histories (i.e., labelled validation loss curves over epochs of training) of DL models that overfit the training data. Our trained time series classifier detects overfitting in a trained DL model by examining the validation loss history (captured as part of the training history). In contrast to existing overfitting detection approaches, our approach does not incur additional resources or costs, as the training history (also known as the learning curve) is a natural byproduct of the training process. Furthermore, our approach (i.e., the trained time series classifier) can be used to prevent overfitting based on the validation losses of recent epochs~(e.g., the last 20 epochs).

Although our approach is trained on a simulated dataset, we evaluate it on a real-world dataset, collected from papers published in top AI venues within the last 5 years. We gathered the training histories from these papers that are explicitly labelled as overfitting or non-overfitting by the authors as the ground truth. The main contributions of this paper are as follows:

\begin{itemize}
   % \item We propose a non-intrusive overfitting detection and prevention approach to enhance the model quality of machine learning systems.
    \item Our approach outperforms the state-of-the-art by at least 5\% in terms of F-score for overfitting detection, achieving an F-score of 0.91.
    \item Our approach has the capability to prevent overfitting at least 32\% earlier than early stopping while maintaining (and often surpassing) the rate of reaching the optimal epoch (i.e., the epoch that yields the best model).
    \item We provide a replication package~\citep{replication_package} containing our trained classifiers and labelled training histories which can be directly used by other researchers.
    %\footnote{\url{https://github.com/OverfitGuard/OverfitGuard}}
\end{itemize}

% Although our approach is trained on a simulated dataset, we evaluate it on a real-world dataset, collected from papers published in top AI venues within the last 5 years. We collected the training histories from these papers that are explicitly labelled as overfitting or non-overfitting by the authors as the ground truth. Our results show that our approach outperforms the state-of-the-art by at least 5\% in terms of F-score for overfitting detection, achieving an F-score of 0.91. In addition, our approach can prevent overfitting from happening at least 32\% earlier than early stopping while maintaining (and often surpassing) the rate of reaching the optimal epoch~(i.e., the epoch that yields the best model). Our work contributes to machine learning software quality assurance by providing a non-intrusive overfitting detection and prevention approach. We provide our trained classifiers and labelled training histories in a replication package\footnote{\url{https://github.com/hisdpo/hisdpo}} which can be directly used by other researchers. 

\textit{Paper organization.} 
This paper is organized as follows. Section~\ref{sec:background} provides background information about our study. Section~\ref{sec:related_work} gives an overview of related work. Section~\ref{sec:existing_approach} introduces existing approaches for detecting and preventing overfitting. Section~\ref{sec:our_approach} describes the design of our study, while Section~\ref{sec:experiment_setup} provides detailed information about our experimental setup. Sections~\ref{sec:rq1} and~\ref{sec:rq2} present the results of our study. Section~\ref{sec:threats} discusses potential threats to the validity of our study. Finally, Section~\ref{sec:conclusion} concludes the paper.

 % Section~\ref{sec:implications} discusses the implications of our findings.
 
\section{Background}\label{sec:background}

This section provides an introduction to the concepts of the training history in DL and time series classification.

\subsection{Leveraging training history in DL}

% https://janvanrijn.github.io/metalearning/content/2022AutoML/tutorial_learning_curves.pdf

Training history, also known as the empirical learning curve~\cite{mohr_learningcurve_2022}, provides valuable insights into a DL model's learning progress and performance throughout the training process. The training history stores a record of metrics during the training process, which are usually recorded in each training iteration or epoch~(as shown in Figure~\ref{fig:examples}). Training loss and validation loss are commonly used metrics in training histories. Typically, a dataset is divided into training, validation, and test sets. While the training loss reflects how well the DL model learns from the training data during the training process, the validation loss is evaluated based on the validation data, which serves as a proxy for evaluating the model's performance on unseen data. After training is completed, the trained DL model's performance is evaluated using the test set that has not been exposed to the model. %The metrics used for evaluating the model are based on the scenario~(e.g., F-score or accuracy) to verify the trained model's capability and generalizability. 

Researchers and developers can identify potential issues such as overfitting or underfitting by analyzing the training histories. For example, overfitting is often observed as an increasing divergence between training loss and validation loss over time~(as illustrated in Figure~\ref{fig:example_overfit}). In this paper, we propose an approach that leverages training history to automatically detect and prevent overfitting in DL models. 

\begin{table}[t]
\centering
% \footnotesize
%\scriptsize
\caption{Studied time series classifiers.}
\label{tab:time_series_classifiers}
\begin{tabularx}{\columnwidth}{l X}
\toprule
Classifier & Description                                               \\ \midrule
KNN-DTW$^*$                & Uses K-Nearest Neighbors~\citep{hand_principles_2007} with Dynamic Time Warping~\citep{ding_dtw_2008} as the distance metric to classify time series data             \\
HMM-GMM                & Uses Hidden Markov Model for modelling time series data and Gaussian Mixture Model as the emissions probability density~\citep{gauvain_hmm_1994, ji_hmm_2006} \\
TSF                    & Uses a random forest~\citep{biau_random_2016} for time series data using an ensemble of time series trees~\citep{deng_tsf_2013}                                       \\
TSBF                   & Time Series Bag-of-Features~\citep{baydogan_tsbf_2013} extracts features based on the bag-of-features approach~\citep{fu_bof_2011} to create a random forest \\
SAX-VSM                & Symbolic Aggregate approXimation~\citep{lin_sax_2007} converts the data into symbolic representations and Vector Space Model~\citep{salton_vsm_1975, peng_tfidf_2014} transforms them into vectors to calculate similarity for classification  \\
BOSSVS                 & Bag-of-SFA Symbols in Vector Space~\citep{schafer_boss_2016} is similar to SAX-VSM but use SFA~\citep{schafer_sfa_2012} to transform the data instead of SAX                       \\ \bottomrule
\multicolumn{2}{l}{\footnotesize{$^*$ KNN-DTW handles variable-length time series data.}} 
\end{tabularx}
\end{table}

\subsection{Time series classification}

Time series data consists of data points recorded over time, with each point being associated with a specific timestamp and its corresponding value. Time series classification is a machine learning task that aims to categorize time series data into predefined classes. In the context of our study, we consider the training history of a DL model as time series data, as the task of identifying whether a DL model is overfitting based on its training history can be framed as a time series classification problem. Since there has been no prior systematic research on time series classifiers specifically designed for training histories, we have selected six classifiers (shown in Table~\ref{tab:time_series_classifiers}) that have been reported as baselines or state-of-the-art in prior studies~\citep{xi2006fast, varol2017early, anami2019comparative, wang2021soil}. These classifiers were chosen due to their demonstrated effectiveness in various time series classification tasks and their potential applicability to the overfitting detection problem in DL.

\section{Related Work}\label{sec:related_work}

%\subsection{Model validation in software systems}

%Model validation in SE serves as a crucial step to ensure the performance, reliability, and generalizability of predictive models. However, the domain has not sufficiently addressed the critical issue of overfitting. To fill the gap in the existing literature, our work focuses on the integration of model validation for both detecting and preventing overfitting.

% While these methods offer valuable insights into model validation, they do not address the issue of overfitting, which is crucial for ensuring model generalizability.

%Moreover, some work has been done in the domain of program repair to specifically address the overfitting issue.However, these methods are often constrained by their focus on program repair and lack general applicability to broader machine learning models.

\subsection{Mitigating overfitting in SE}

Overfitting poses a significant risk to the trustworthiness of software systems and the research studies that employ DL models. SE researchers typically use either overfitting detection or prevention methods to mitigate the problem of overfitting. Among the overfitting prevention methods, dropout is the most commonly adopted approach~\citep{yang_survey_2022, watson_survey_2022}. Researchers have used dropout in various domains such as code generation~\citep{liu_dropout_2020, liu_dropout2_2020}, logging locations recommendation~\citep{li_dropout_2020}, and comment completion~\citep{ciurumelea_dropout_2020, wei_dropout_2021}. Regularization is another prominent overfitting prevention strategy that has been used in SE~\citep{barbez_regularization_2019, zampetti_regu_2020, huang_regu_2020, bowen_regu_2016}, which requires adding another layer to the model structure as well. For example, \citet{zampetti_regu_2020} employed L2-norm regularization in training CNN and RNN models to manage self-admitted technical debt in source code.

Early stopping is another frequently used technique to prevent overfitting during the training process of DL models~\citep{hoang_es_2019, choetkiertikul_es_2019, shi_early_2020}. For example, \citet{shi_early_2020} utilized early stopping when training a deep Siamese network to identify hidden feature requests posted in chat messages by developers. Other techniques like data augmentation~\citep{mahadi_dataaug_2020, fakhoury_dataaug_2018, bao_dataaug_2020} and data balancing~\citep{ren_data_2019, tufano_data_2019} are also employed to address overfitting. For instance, \citet{bao_dataaug_2020} developed a CNN-based image classification model to filter out non-code and noisy-code frames from programming screencasts. To enhance training data diversity, they employed data augmentation techniques such as rotation, scaling, translation, and shearing.

In terms of overfitting detection approaches, \citet{zhang_pv_2019} propose the perturbation validation~(PV) assessment to determine whether a DL model fits the training data properly (i.e., ensure that it is neither overfitting nor underfit). Alternatively, some detection approaches check the hypothesis that the trained DL model and the data are independent. For example, \citet{werpachowski_detecting_2019} check the hypothesis by comparing the test error with the estimated test error based on adversarial examples of the test set.

Another popular approach towards detecting overfiting is to use model validation approaches. \citet{tanti_validation_2017} evaluated 12 model validation techniques specifically for defect prediction models and concluded that out-of-sample bootstrap emerges as the least biased and most stable technique that helps detect overfit. \citet{damiani_certified_2020} introduced a framework that validates DL models against desired non-functional properties and statistically monitors the model output. \citet{straub_validation_2021} extended this by utilizing randomly generated expert networks for model validation that focuses on performance characteristics.

Other than these approaches, \citet{smith_test_2015} discussed the role of human factors in overfitting and provided a comparative analysis between automated patches and human-written fixes. They reported that overfitting is not solely a machine-induced problem and suggested focusing on the contributing factors like test suite coverage and requirements-based testing. \citet{xin_test_2017} proposed a classification technique that differentiates overfitting patches based on semantic differences. \citet{nilizadeh_test_2021} utilized formal verification to evaluate the degree of overfitting and identified the challenges posed by program complexity and numeric issues.

To the best of our knowledge, both the overfitting prevention and detection methods used in both SE studies and in practice fall prey to several key concerns. The overfitting prevention methods, typically require significant expertise to execute correctly and are intrusive (for instance, they may require one to modify the data or the model structure). Overfitting detection approaches typically require retraining of the DL model multiple times, which may be very costly in practice~\cite{fu2017easy} and simple methods like early stop may stop the training of DL model sub optimally. Our work addresses these gaps by introducing a history-based approach that serves the dual purpose of detecting and preventing overfitting in a non-intrusive manner without any need for DL model retraining.

\begin{figure*}[t]
     \centering
     \begin{subfigure}[b]{0.8\columnwidth}
        \centering
        \includegraphics[width=\columnwidth]{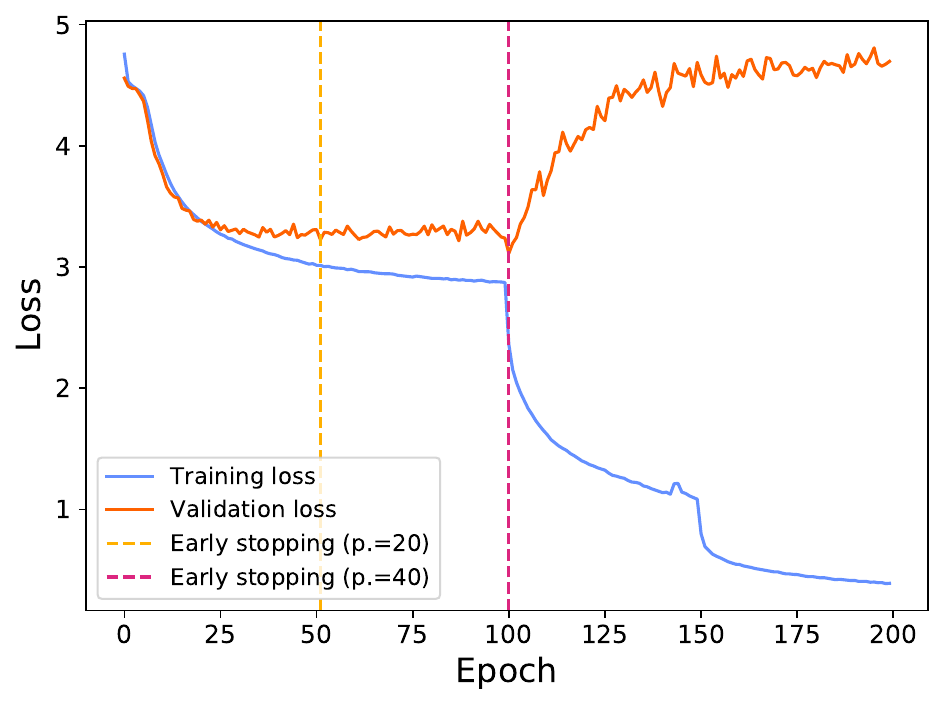}
        \caption{Examples of early stopping with the patience of 20 and 40 epochs~(stop at the \#71 epoch and \#140 epoch).}
        \label{fig:early_stop_example}
    \end{subfigure}
    \hspace{0.1\columnwidth}% \hfill
    \begin{subfigure}[b]{0.8\columnwidth}
         \centering
         \includegraphics[width=\columnwidth]{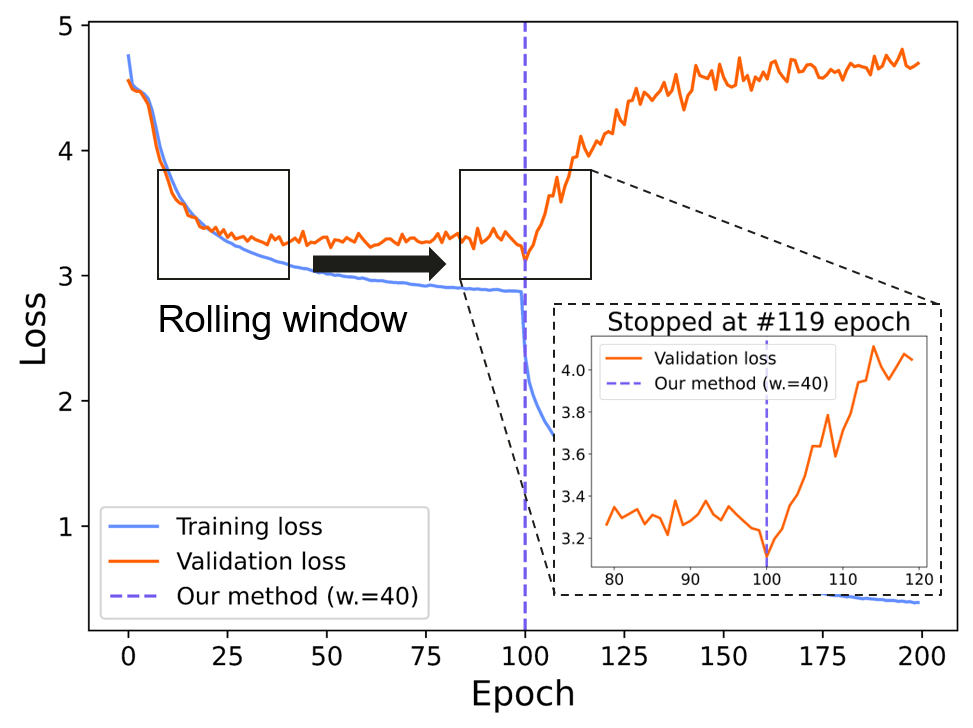}
         \caption{An demonstration of our overfitting prevention approach with a rolling window.}
         \label{fig:tsc_example}
    \end{subfigure}
    \caption{Early stopping and our approach for overfitting prevention.}
    \label{fig:prevention_example}
\end{figure*}

\subsection{Leveraging training history to improve DL software quality}

While the machine learning community has leveraged training histories~(i.e., learning curves) for different tasks, the SE community has seldom used training history to enhance software quality. \citet{mohr_learningcurve_2022} conduct a survey on approaches based on learning curves for decision-making in DL domains, such as data acquisition, early stopping, and model selection. They also propose an approach called Learning Curve Cross-Validation (LCCV)~\citep{mohr2021towards} that iteratively increases the number of training examples used for training to select the best model from the candidates. Training histories can also be used to evaluate the trained classifiers: \citet{rijn2015fast} propose an approach that recommends classifiers for a given dataset based on training histories using loss time curves. Moreover, \citet{hoiem_learncurve_2021} investigate the use of training histories to evaluate design choices for DL models, such as pretraining, architecture, and data augmentation. 

In this paper, we propose \OverfitGuard, an approach that utilizes time-series classifiers to detect and prevent overfitting by analyzing the training history of DL models. Our approach aims to enhance the software quality of DL systems by improving the quality of the DL models themselves. \OverfitGuard is one of the first approaches that leverages training history to enhance the software quality of DL systems. A related work by \citet{tokui_neurecover_2022} introduces an approach called NeuRecover which also leverages training history to improve the quality of DL models, particularly for safety-critical applications. NeuRecover identifies the model parameters that need to be modified (i.e., repaired) by analyzing the training history to address specific failure types. Our approach shares similarities with NeuRecover, but focuses on the detection and prevention of overfitting, which is important for maintaining the software quality of DL systems.

% \citet{giray_se4ml_2021} conduct a survey about
% 'Dealing with new types of quality attributes'
% ' Testing \& Quality is the most focused knowledge area by academia and industry'
% `researchers collaborated the most within the scope of testing and quality.`

% \citet{zhang_mltest_2022} ML testing. 

% \citet{lo_fedlearn_2021} Federated Machine Learning, analyze 231 primary studies to identify the state-of-the-art in federated learning and explore how to develop federated learning systems.

% \citet{paleyes_deploy_2022} discuss the challenges associated with deploying machine learning models in production systems

% \citet{amershi_se4ml_2019} A SE for ML case study in Microsoft

% \citet{masuda_sq4ml_2018}, sq for ML. highlights specific quality concerns associated with ML software, 

% In this paper, we propose \OverfitGuard. We are one of the first researchers that leverages training history to enhance the software quality of ML systems. One related work is NeuRecover.

% In addition, \citet{zhang_mlse_2003} provide suggestions for applying ML techniques to SE tasks. They also outline previous work in this subject and provide an overview of the applicability of various ML algorithms in the SE domain.

% \citet{ghotra_revisit_2015} investigate the impact of classification techniques on the performance of defect prediction models and discover that the choice of classification technique has little influence on model performance. 
\section{Existing approach}\label{sec:existing_approach}

In this section, we introduce the existing approaches that we use as baseline approaches to compare our proposed approach for overfitting detection and prevention in DL models.

\subsection{Overfitting detection}~\label{sec:exsiting_detect}

% \begin{figure*}[t]
% \begin{center}
% \includegraphics[width=0.8\textwidth]{existing_overfitting_detection}
% \end{center}
% \caption{Correlation-based approaches and PV measurement for overfitting detection.}
% \label{fig:existing_detection}
% \end{figure*}

\textbf{Correlation-based approaches.} 
One approach for detecting overfitting in DL models is to compute the correlation between the training and validation loss. \citet{kronberger_overfitting_2011} propose computing the non-parametric Spearman's rank correlation coefficient~\cite{spearman_1987} between training and validation fitness in symbolic regression models to detect overfitting. Similar to our approach, correlation-based approaches also detect overfitting based on the training history. Therefore, we select them as a baseline for comparison. In this study, we calculate the correlation metrics between the training and validation loss to detect overfitting in DL models. The underlying principle is straightforward: when no overfitting occurs, the training and validation losses should be correlated, whereas weak correlation implies overfitting. 

We determine the presence of overfitting by comparing the computed correlation-based metric between training and validation losses with a predetermined threshold value~(determined in Section~\ref{sec:exps}). We use three different correlation metrics: Spearman, Pearson~\citep{hauke_correlation_2011}, and time-lagged Pearson correlation coefficients. Both Spearman and Pearson correlation coefficients are calculated since we do not know whether the relationship between training and validation loss is linear or not. In addition, we compute the Pearson correlation coefficient between a 5-epoch lagged version of the training loss and the validation loss. This approach is inspired by autocorrelation~\citep{brockwell_timeseries_2002}, which measures the correlation between a time series data and a time-lagged version of itself.

% \textbf{Perturbation validation measurement.} 
% Another approach for detecting overfitting is to use perturbation validation (PV) assessment. \citet{zhang_pv_2019} propose the perturbation validation~(PV) assessment to determine whether a model fits the training data properly (i.e., ensure that it is neither overfitting nor underfit). As shown in Figure~\ref{fig:existing_detection}, we inject three levels of noise~(0.1, 0.2, and 0.3), as suggested by the original paper~\citep{zhang_pv_2019}, into the labels in the training set and retrain the model. We repeat this process for each noise level, collect the training history, and compute the PV measurement. The PV measurement reveals the rate of accuracy decline in response to injected noise, indicating if overfitting is present. The idea behind this approach is that overfit or underfit models would lose accuracy more slowly when trained using the noise-injected training set than optimally-fitted models. Since the calculated PV measurement is only one value, we compare it with a threshold to determine if there is overfitting. We describe how we determine the threshold in Section~\ref{sec:exps}. 

\subsection{Overfitting prevention} \label{sec:existing_prevent}

\textbf{Early stopping.} 
One widely used approach for preventing overfitting is early stopping, which stops training when there is no improvement in a fixed number of epochs (called patience) and returns the epoch which has the lowest validation loss. We choose the widely used TensorFlow implementation,\footnote{\url{https://tensorflow.org/api_docs/python/tf/keras/callbacks/EarlyStopping}} which is also used by PyTorch Ignite.\footnote{\url{https://pytorch.org/ignite/generated/ignite.handlers.early_stopping.EarlyStopping}} As shown in Figure~\ref{fig:early_stop_example}, early stopping with a patience parameter of 20 epochs stops at the \#70 epoch and returns the \#50 with the lowest validation loss since no improvement occurs between epochs \#50 and \#70. Furthermore, early stopping with larger patience values (e.g., 40 epochs) stops later~(at the \#140 epoch) but with a lower loss~(at the \#100 epoch).

\textbf{Early stopping based on smoothed validation loss curves.}
An alternate version of early stopping inspects the moving average of the smoothed validation loss curves~\citep{shumway_time_2017, molugaram_chapter_2017} to decide when to stop the training process. After stopping the training process, this approach returns the best epoch which has the lowest validation loss (not the smoothed value).

\begin{figure*}[t]
\begin{center}
\includegraphics[width=0.8\textwidth]{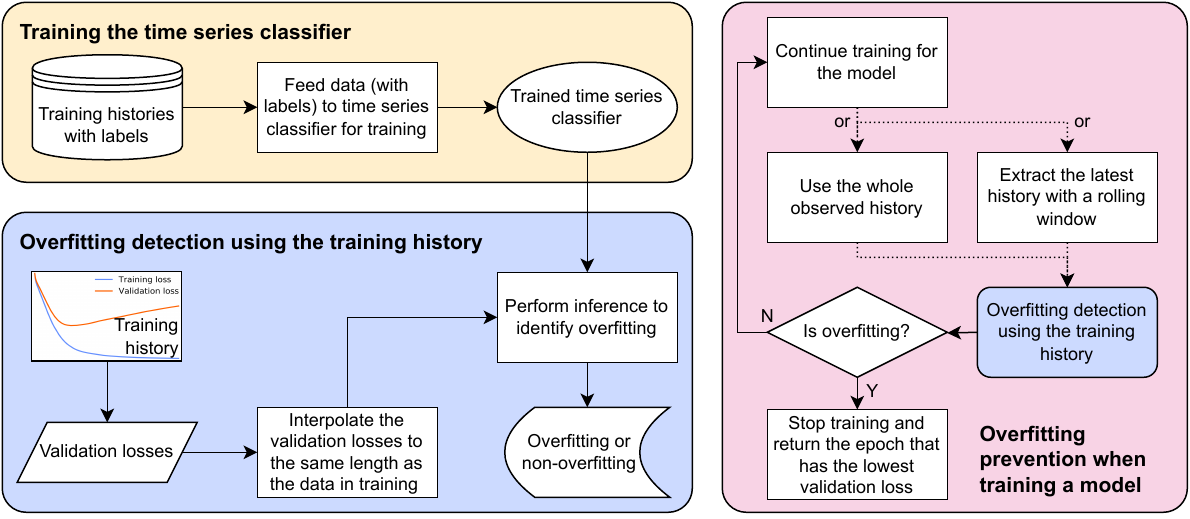}
\end{center}
\caption{Our approach for overfitting detection and prevention.}
\label{fig:our_method}
\end{figure*}

\section{Our approach}\label{sec:our_approach}

Figure~\ref{fig:our_method} shows an overview of our proposed approach. Our approach uses a time series classifier to detect and prevent overfitting. Table~\ref{tab:time_series_classifiers} lists the studied time series classifiers. First, we collect a simulated dataset (more details on how we collect the data in Section~\ref{sec:experiment_setup}) that contains training histories (i.e., training and validation loss curves, however, we only use the validation loss curves in our approach) with labels indicating whether overfitting occurs in order to train our time series classifier. Second, we train and evaluate each studied time series classifier on all of the training histories of the simulated dataset. Finally, we use the trained time series classifier to perform both overfitting detection and prevention as follows.

% \begin{figure}[b]
% \begin{center}
% \centerline{\includegraphics[width=\columnwidth]{interpolate}}
% \caption{An example of linear interpolation.}
% \label{fig:interpolate}
% \end{center}
% \end{figure}

\textbf{Overfitting detection.} To detect overfitting in a trained DL model, we first collect its validation losses over the training epochs. We feed this loss to our trained time series classifier to detect whether there is overfitting. However, we cannot directly feed these validation losses to our classifier, since the length of the validation losses might not be of the same length as that of the data used to train these time series classifiers. All the studied time series classifiers, with the exception of KNN-DTW, expect the length of the inputs used for training and inference to be the same. Therefore, we first linearly interpolate the validation losses of the DL model for which we need to detect overfit to the same length as the training histories used to train the studied time series classifiers. We feed the interpolated validation losses to our trained time series classifier and perform inference to determine if the DL model is overfit. %Figure~\ref{fig:interpolate} shows how the linear interpolation process works; if we only have validation losses over 8 epochs and our time series classifier was trained over 80 epoch validation loss values, we interpolate the 8 epoch losses to 80 so that we can feed it to the trained time series classifier. 

\textbf{Overfitting prevention.} To prevent overfitting, we feed the training history (i.e., validation loss curve) of a DL model that is being trained to our trained time series classifier during the training process. The history is fed for inference in two different ways: (1)~as a rolling window: we extract the latest history in a fixed window size~(e.g., the latest 20 epochs), and (2)~as the whole observed history (from the first to the latest epochs). Our time series classifier detects if in the fed history overfitting occurs. Similar to overfitting detection, we linearly interpolate the data before feeding it into our model. If there is no overfit occurring, we continue the training and repeat the above procedure until the DL model has finished training. For the rolling window, we move the window by a fixed step size~(as shown in Figure~\ref{fig:tsc_example}) and make another prediction. If our model detects the presence of overfitting in the fed history, we return the lowest validation loss in the observed epochs as the best epoch.

\section{Experimental setup}\label{sec:experiment_setup}

\begin{figure}[t]
\begin{center}
\includegraphics[width=\columnwidth]{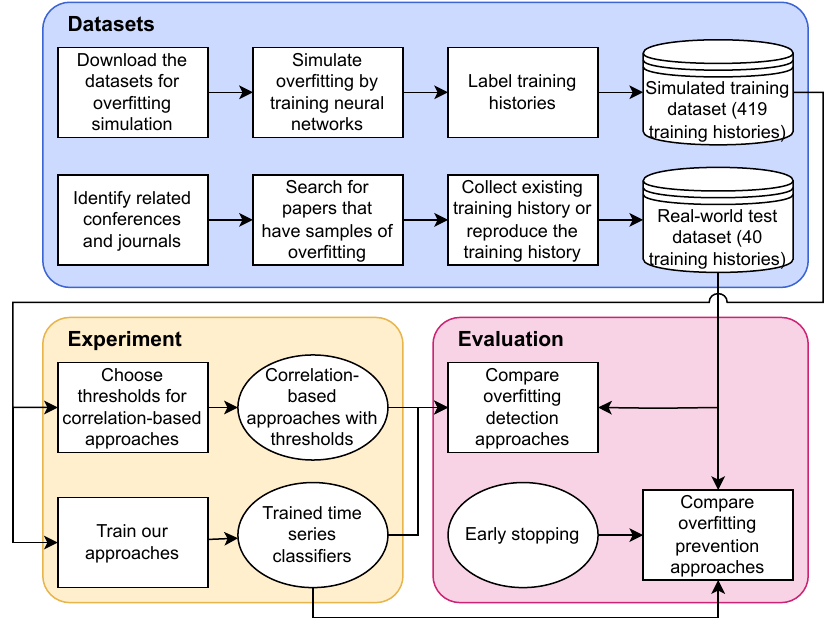}
\end{center}
\caption{Overview of the experimental setup.}
\label{fig:experiment_overview}
\end{figure}

In this section, we introduce the datasets for training and evaluating the studied overfitting detection and prevention approaches, the experiments of our study, and the evaluation metrics for the studied approaches. Figure~\ref{fig:experiment_overview} shows an overview of the experimental setup.

\subsection{Environment setting}

We conducted the experiments on an Ubuntu 20.04 operating system with a Linux kernel version of 5.15.0, utilizing Python 3.8 and TensorFlow 2.9.0. The hardware configuration for the experiments is detailed below:

\begin{itemize}
    \item NVIDIA RTX 3090 GPU with 24 GB memory
    \item CUDA version 10.1.243
    \item cuDNN version 7.6.5
    \item Intel(R) Core(TM) i9-11900K CPU with a clock speed of 3.50 GHz
    \item 64 GB of RAM
\end{itemize}

\begin{table}[t]
\centering	
\caption{Information about datasets used to simulate overfitting.}
\label{tab:proben1}
\begin{tabular}{llrrr}
\toprule
Dataset  & Type & \#In & \#Out & \#Examples \\ \midrule
building & regression     &  14 &  3  & 4,208  \\
cancer   & classification &   9 &  2  &   699  \\
card     & classification &  51 &  2  &   690  \\
diabetes & classification &   8 &  2  &   768  \\
flare    & regression     &  24 &  3  & 1,066  \\
gene     & classification & 120 &  3  & 3,175  \\
glass    & classification &   9 &  6  &   214  \\
heart    & classification &  35 &  2  &   920  \\
hearta   & regression     &  35 &  1  &   920  \\
horse    & classification &  58 &  3  &   364  \\
soybean  & classification &  82 & 19 &   683  \\
thyroid  & classification &  21 &  3  & 7,200  \\ \bottomrule
\end{tabular}
\end{table}

\subsection{Simulated training dataset}\label{sec:simulated_dataset}

We create a simulated dataset containing training histories with labels to determine a threshold for correlation-based approaches (see Section~\ref{sec:existing_approach}) and to train our proposed method as described in Section~\ref{sec:our_approach}. We create this simulated dataset by training neural networks of varying model complexities to produce overfitting and non-overfitting samples. The process is as follows:

\textit{Step 1 -- Download the datasets for overfitting simulation.} We download 12 datasets representing real-world problems from the Proben1~\citep{proben1} benchmark set for training neural networks. These datasets were used by \citet{prechelt_early_2012} to simulate training histories for studying early stopping. We choose these datasets over using SE datasets for two reasons: First, since we use the methodology used by~\citet{prechelt_early_2012} to simulate overfitting, we chose to stick with the datasets that they used. Second, irrespective of the domain of the dataset, we assert that how the phenomenon of overfit is represented by training and validation histories will remain the same. Table~\ref{tab:proben1} provides information about these datasets, which include 3 datasets for regression tasks and 9 datasets for classification tasks. All of these datasets~(except the ``building'' one) were originally collected from the UCI machine learning repository~\citep{blake_uci_1998}, which has been widely used in DL research~\citep{kuleshov_uci_2018, sajeev_uci_2019, shi_uci_2021, gadde_uci_2021}. Each dataset is pre-partitioned into training, validation, and test sets (50\%, 25\%, and 25\% of the data, respectively) and partitioned three times to generate three distinct permutations, resulting in a total of 36 datasets from Proben1.

\textit{Step 2 -- Simulate overfitting by training neural networks.} We train neural networks (NNs) with various architectures on the collected 36 datasets. We do so to vary the model complexity which in turn increases the chance of producing an overfitted DL model, following the methodology of \citet{prechelt_early_2012} used in their study. The input/output layer of each NN contains the same number of nodes as the number of input/output coefficients in the respective datasets~(see Table~\ref{tab:proben1}) and rectified linear units~(ReLUs) are used for all hidden layers. The structures of the NNs are as follows: (1)~6 one-hidden-layer NNs with 2, 4, 8, 16, 24, or 32 hidden nodes, and (2)~6 two-hidden-layer NNs with hidden nodes (represented as first layer hidden nodes + second layer hidden nodes) of 2+2, 4+2, 4+4, 8+4, 8+8, 16+8. We use the mean square error~(MSE) as the loss function for regression problems, and cross entropy as the loss function for classification problems. All problems employ stochastic gradient descent (SGD) as the optimizer. To increase the likelihood of overfitting, we train these 12 neural network architectures on each of the 36 datasets for 1,000 epochs, producing 432 training histories.

\textit{Step 3 -- Label training histories.} 
%To train our proposed approach and the correlation-based approaches, labelled training histories are required. However, due to the abstract nature of overfitting definitions in the literature~\citep{anderson_overfit_2004, hawkins2004problem, ying_overfit_2019, systematic_bejani_2021}, developing a heuristic rule from manually specified criteria is challenging. We demonstrate in Appendix~\ref{appendix:auto_label} that a heuristic approach for labelling training histories performs poorly on the simulated dataset. As a result, we decided to label the training histories manually by human experts.
To ensure the robustness of our manual labelling process, we follow the approach outlined by \citet{ding_label_2020}. The first and second authors of this paper independently labelled the 432 data points as either ``overfit'', ``non-overfit'' or ``uncertain'' and discussed the results. In the first discussion round, the authors reached a 95\% agreement~(410 data points), with both authors labelling 10 data points as "uncertain" and subsequently eliminating them. In the second round, the authors discussed the remaining 22 disagreements. Following the discussion, we eliminated 3 data points~(labelled ``uncertain'' by both authors) and agreed on the labels for the remaining 19 data points. The final dataset consists of 44 overfit and 375 non-overfit training histories. We share the labelled training histories in our replication package for other researchers to reuse.

\subsection{Real-world test dataset}~\label{sec:real_world_dataset}

To evaluate our approach using real-world data, we conducted a survey of papers from conferences and journals to gather examples of overfit and non-overfit DL models. %We do so since these examples unlike the manual labelling that we did in the previous step are not subject to labeller bias and are widely accepted by the research community to indeed be representative of training histories pertaining to overfit and non-overfit DL models.

\textit{Step 1 -- Identify related conferences and journals.} 
We identify related conferences and journals based on the Computing Research and
Education Association of Australasia~(CORE\footnote{\url{https://www.core.edu.au}}) and China Computer Federation~(CCF\footnote{\url{https://ccf.atom.im/}}) ranking systems. Under the CCF A rank, we have 7 conferences and 4 journals in the ``Artificial Intelligence'' field. Under the CORE A* rank, we have 16 conferences in the ``machine learning'' and ``artificial intelligence'' fields and 12 journals in the ``artificial intelligence and image processing'' field. After merging the results and accounting for overlaps between the two ranking systems, we obtained a final list of 17 conferences and 12 journals.

We collected our real-world data in the artificial intelligence and machine learning domain as opposed to SE domain, because SE studies typically do not report the training histories of the overfit DL models and we required community accepted examples of the training histories of both overfit and non-overfit DL models. 

\textit{Step 2 -- Search for papers that have samples of overfitting.} 
We found 33 full papers~(see the replication package~\citep{replication_package}) containing the keyword "overfit" (including variations such as "overfitting") in the title that were published in the selected conferences and journals in the last 5 years. Five of these papers provided samples of overfitting: P2 - \citet{chatterjee_circuit_2020}; P4 - \citet{chen_robust_2021} P13 - \citet{Kim_understanding_2021}; P17 - \citet{rice_overfitting_2020} and P23 - \citet{singla_low_2021}. Table~\ref{tab:survey_examples} lists the papers and the number of collected samples of overfitting~(some of them also provide samples of non-overfitting). 

% \input{tabs/surveyed_papers}

% Moved to appendix
\begin{table}[t]
\caption{Information about collected samples from surveyed papers.}
\label{tab:survey_examples}
\scriptsize
\begin{tabularx}{\linewidth}{ l X r r}

\toprule
Paper & Labels for the training history in the manuscript & \#Overfit & \#Non-overfit \\ \midrule
P2      &
% P2~\citet{chatterjee_circuit_2020}      &
``[...] the validation accuracy of nn-random is 9.73\% (i.e., close to chance) confirming that it is horribly overfit''
&   2        &  0            \\
P4      &
% P4~\citet{chen_robust_2021}      &
``We first observe that the robust overfitting prevails in all Baseline cases''
\newline
``Our methods effectively mitigates the robust overfitting''
&   3        &  3             \\
P13      &
% P13~\citet{Kim_understanding_2021}      &
``Figure 4 [...] that is, catastrophic overfitting occurs.''
\newline
``Figure 6 shows that the proposed method also successfully prevents catastrophic overfitting [...]''
&  N/A$^*$        & N/A$^*$             \\ 
P17      &
% P17~\citet{rice_overfitting_2020}      &
% ``Table 3 [...] We observe robust overfitting to occur across all experiments''
% \newline
``Figure 24 [...] We see clear robust overfitting for the smaller two options in $\lambda$, and find no overfitting but highly regularized models for the larger two options [...]''
&  20        &  4              \\
P23      &   
% P23~\citet{singla_low_2021}      &      
``These results therefore validate our claim that low curvature activations reduce robust overfitting''
&   4        &  4             \\\bottomrule
\multicolumn{4}{l}{\footnotesize{$^*$ Cannot reproduce the same results as the paper.}} 
\end{tabularx}
\end{table}

\textit{Step 3 -- Collect existing training history or reproduce the training history.}
Paper P17 shared the training history, making its replication straightforward. We replicated the other papers that provide overfitting samples to collect the training histories of these samples. We executed the code from the papers with available replication packages (P4, P13, and P23) to generate their training histories. However, we could not replicate the results for paper P13. For paper P2, which did not provide a replication package, we followed the methodology to replicate the results and training history. In total, we collected 29 training histories of overfit DL models and 11 of non-overfit DL models~(refer to Table~\ref{tab:survey_examples} for details). 
\subsection{Experiments}\label{sec:exps}

\textbf{Overfitting detection.} 
We trained the time series classifiers based on the simulated dataset. For each classifier, a grid search with 3-fold cross-validation was performed to tune the hyperparameters based on the simulated dataset. Once the optimal hyperparameters were identified, we proceeded to train each time series classifier using all training histories and labels from the simulated dataset and saved the trained classifier for further use. For correlation-based approaches, we also performed a grid search based on the simulated dataset to select the optimal thresholds (ranging from -1 to 1) that yielded the best F-score.% Furthermore, we conducted experiments using the intrusive PV approach for comparison purposes with the other overfitting detection approaches described above. To choose the appropriate threshold, we carried out a grid search for the thresholds between -1 and 1 based on the F-score on the simulated dataset.

\textbf{Overfitting prevention.}
We reused the trained time series classifiers from the previous step to perform inference during the training process to prevent overfitting. Since the validation loss curve is generally applicable to both classification and regression tasks, we used it for overfitting prevention. We applied our approach to the trained DL models in every 10 epochs~(i.e., the step size), with varying rolling window sizes of 20, 40, 60, 80, and 100 epochs. We used early stopping based on the validation loss and set the patience values to range from 5 to 115 epochs. We also applied early stopping based on smoothed validation loss curves generated by a 10-epoch moving average~\citep{shumway_time_2017, molugaram_chapter_2017}. 
% While overfitting prevention based on the validation loss curve is generally applicable for both classification and regression tasks, the classification error (i.e., zero-one loss) curve might be used specifically for classification tasks to prevent overfitting. Experiments of overfitting prevention approaches based on the zero-one loss are provided in Appendix~\ref{appendix:acc_curve}.

\subsection{Evaluation}\label{sec:evaluation}

\textbf{Evaluation metrics for overfitting detection.} 
To evaluate the classification performance of the overfitting detection approaches, we computed the \textbf{precision}, \textbf{recall}, and \textbf{F-score} for overfitting and non-overfitting samples in the real-world test dataset. In addition, we calculated the \textbf{average F-score} to directly compare the classification performance of the studied approaches. To evaluate the time cost associated with training and using the studied approaches, we report the \textbf{training time} (in seconds) for each approach on the simulated dataset and the \textbf{inference time} (in milliseconds) for the real-world dataset. 

\textbf{Evaluation metrics for overfitting prevention.} 
Ideally, an overfitting prevention approach returns the optimal epoch~(i.e., the epoch that yields the best predictive performance for the DL model on the validation set) and stops the training process as early as possible. We define the \textbf{optimal rate} of an overfitting prevention approach as the percentage of cases where the optimal epoch is successfully identified. To assess the speed of the approach, we introduce the \textbf{delay} metric, which represents the epoch difference between the stopped epoch and the best epoch. For example, a delay of 10 epochs occurs if the prevention approach stops at the $123^{th}$ epoch while the $113^{th}$ epoch is the best one. In addition, we report the DL model's \textbf{accuracy} on the validation set when the training process is stopped by the overfitting prevention approach.

\begin{table*}[t]
\begin{center}
% \begin{small}
% \begin{sc}
\caption{Results of the overfitting detection approaches on the simulated dataset (CV F-S: F-score of cross-validation) and real-world dataset (Prec: precision; Rec: recall; F-s: F-score; Avg F-s: average F-score), and the time cost of training the studied approaches on the simulated dataset and performing inference on the real-world dataset (per sample).}
\label{tab:res_test_set}
\begin{tabular}{ll|rrr|rrr|r|r|rr}
\toprule
\multicolumn{2}{l|}{\multirow{2}{*}{Detection approach}} &
  \multicolumn{3}{c|}{Non-overfitting} &
  \multicolumn{3}{c|}{Overfitting} &
  \multicolumn{1}{r|}{\multirow{2}{*}{\begin{tabular}[c]{@{}r@{}}Avg \\ F-s\end{tabular}}} &
  \multicolumn{1}{r|}{\multirow{2}{*}{\begin{tabular}[c]{@{}r@{}}CV  \\ F-s\end{tabular}}} &
  \multicolumn{1}{l}{\multirow{2}{*}{\begin{tabular}[c]{@{}l@{}}Training\\ time (s)\end{tabular}}} &
  \multicolumn{1}{l}{\multirow{2}{*}{\begin{tabular}[c]{@{}l@{}}Inference\\ time (ms)\end{tabular}}} \\
\multicolumn{2}{l|}{} &
  Prec &
  Rec &
  F-s &
  \multicolumn{1}{r}{Prec} &
  \multicolumn{1}{r}{Rec} &
  \multicolumn{1}{r|}{F-s} &
  \multicolumn{1}{r|}{} &
  \multicolumn{1}{r|}{} &
  \multicolumn{1}{l}{} &
  \multicolumn{1}{l}{} \\ \midrule
\multirow{3}{*}{\begin{tabular}[c]{@{}l@{}}Corr.\\ based\end{tabular}} &
  Spearman &
  0.71 &
  0.91 &
  0.80 &
  0.96 &
  0.86 &
  0.91 &
  \textbf{0.86} &
  0.95 &
  2.461 &
  0.908 \\
 &
  Pearson &
  0.78 &
  0.64 &
  0.70 &
  0.87 &
  0.93 &
  0.90 &
  0.80 &
  0.92 &
  0.222 &
  0.025 \\
 &
  Autocorr &
  0.80 &
  0.73 &
  0.76 &
  0.90 &
  0.93 &
  0.92 &
  0.84 &
  0.80 &
  0.233 &
  0.026 \\ \midrule
\multirow{6}{*}{\textbf{\begin{tabular}[c]{@{}l@{}}Time \\ series\\ (ours)\end{tabular}}} &
  KNN-DTW &
  0.79 &
  1.00 &
  0.88 &
  1.00 &
  0.90 &
  0.95 &
  \textbf{0.91} &
  0.97 &
  0.001 &
  180.512 \\
 &
  HMM-GMM &
  0.30 &
  0.27 &
  0.28 &
  0.73 &
  0.76 &
  0.75 &
  0.52 &
  0.59 &
  99.751 &
  17.750 \\
 &
  TSF &
  0.77 &
  0.91 &
  0.83 &
  0.96 &
  0.90 &
  0.93 &
  0.88 &
  0.99 &
  0.311 &
  17.209 \\
 &
  TSBF &
  0.79 &
  1.00 &
  0.88 &
  1.00 &
  0.90 &
  0.95 &
  \textbf{0.91} &
  0.99 &
  0.301 &
  31.683 \\
 &
  BOSSVS &
  0.46 &
  0.91 &
  0.61 &
  0.94 &
  0.59 &
  0.72 &
  0.67 &
  1.00 &
  1.877 &
  19.342 \\
 &
  SAX-VSM &
  0.83 &
  0.91 &
  0.87 &
  0.96 &
  0.93 &
  0.95 &
  \textbf{0.91} &
  0.96 &
  0.912 &
  17.474 \\ \bottomrule
% \multicolumn{12}{p{13cm}}{\footnotesize{$^*$ We report the F-score of correlation-based approaches (no cross-validation) on the whole simulated dataset.}} 
\end{tabular}
% \end{sc}
% \end{small}
\end{center}
\end{table*}

\section{RQ1: \rqone}\label{sec:rq1}

\motivation
Overfitting detection is an important task in DL models since it helps in identifying whether a DL model has learned to perform well on training data but fails to generalize on unseen data. Accurate overfitting detection can assist researchers and developers in making informed decisions regarding model selection, hyperparameter tuning, and other model performance improvements. This research question investigates the performance of our proposed approach for detecting overfitting in trained DL models and compares it with existing correlation-based approaches.

\approach
We use the evaluation metrics introduced in Section~\ref{sec:evaluation} to compare our approach with baseline approaches based on the real-world test dataset. Furthermore, we record the F-score obtained from the 3-fold cross-validation~(CV) for our approach based on the simulated training dataset to further analyze the performance of our approach. Since we use the entire simulated training dataset to determine the thresholds (without CV) for correlation-based approaches, we report the F-score for correlation-based approaches based on the whole simulated training dataset. 

\results
\textit{Overfit DL models can be detected by inspecting the training history, and our approach using time series classifiers demonstrates better classification performance than the correlation-based approaches for overfitting detection.} Table~\ref{tab:res_test_set} shows that our approach using KNN-DTW, TSBF, and SAX-VSM generalizes well from the simulated dataset to the real-world dataset with the best F-score~(0.91), followed by TSF which outperforms the baseline approaches as well. In contrast, HMM-GMM performs poorly on both the simulated training and real-world test datasets. One possible explanation is that the extracted state models (via HMM) of the training histories do not follow a Gaussian probability distribution. Our approach with BOSSVS correctly identifies all the data in the simulated dataset but performs poorly on the real-world dataset. One reason could be that the extracted bag-of-SFA symbols~(BOSS) from the simulated dataset does not generalize to the real-world dataset. In addition, we note that the investigated correlation-based approaches perform reasonably well, with F-scores greater than 0.8. However, our approach outperforms the correlation-based overfitting detection approach by at least 5\% on the studied real-world dataset.

\textit{The studied time series classifiers are more computationally intensive than correlation-based approaches for inference, yet they are still useful in practice.} As shown in Table~\ref{tab:res_test_set}, our approach requires more time for performing inference than the correlation-based approaches. For instance, TSF has the fastest inference time among the classifiers but is around 20 times slower than the Spearman correlation-based approach and around 700 times slower than the other two correlation-based approaches. However, the speed of our approach is not prohibitive in practice since overfitting detection is only executed once after the training is complete. It is also useful to note that the training times of the time series classifiers in our approach are not excessive. For instance, the training times of TSF and TSBF are around 300 milliseconds, and KNN-DTW, our best-performing time series classifier, can finish training in 1 millisecond. However, KNN-DTW requires the longest time for inference which is around 180 milliseconds for a training history. A fast version of DTW~\citep{salvador_fastdtw_2007} with a time complexity of $O(n)$ is used in experiments, but using KNN with DTW is still computationally intensive.

\hypobox{
	\emp{RQ1 Takeaway:} 
Our proposed approach demonstrates better classification performance than correlation-based approaches for detecting overfitting in DL models. Despite the higher computational cost of the time series classifiers used in our approach, their training time and inference time are still practical.
}
% \end{Summary}

\begin{table}[t]
\caption{The optimal rate, median delay, and average accuracy of our overfitting prevention approaches using the whole observed history.}
\label{tab:whole_his_prevention}
\begin{center}
\begin{tabular}{lrrr}
    \toprule
    \multirow{2}{*}{Classifier} & \multirow{2}{*}{\begin{tabular}[c]{@{}r@{}}Optimal\\ rate\end{tabular}} & \multirow{2}{*}{\begin{tabular}[c]{@{}r@{}}Median\\ delay\end{tabular}} & \multirow{2}{*}{\begin{tabular}[c]{@{}r@{}}Average\\ accuracy\end{tabular}} \\
            &      &      \\ \midrule
    KNN-DTW & \textbf{0.95} & \textbf{43.5} & \textbf{0.42} \\
    HMM-GMM & 0.18 & 0.0  & 0.36 \\
    TSF     & 0.90 & 35.0 & 0.42 \\
    TSBF    & 0.83 & 31.0 & 0.41 \\
    BOSSVS  & 0.65 & 21.0 & 0.41 \\
    SAX-VSM & 0.33 & 10.0 & 0.38 \\ \bottomrule
\end{tabular}
\end{center}
\end{table}

\section{RQ2: \rqtwo}\label{sec:rq2}

\motivation
Another critical part of developing trustworthy and stable DL models is preventing overfitting. An effective overfitting prevention approach allows DL models to generalize better on unseen data while minimizing both training resources and computational costs. This research question evaluates the performance of our proposed approach for preventing overfitting during the training process compared with the frequently used early stopping approach.

\approach
We assess our overfitting prevention approach against the early stopping method (both with and without smoothing loss curves) using the metrics introduced in Section~\ref{sec:evaluation}. To study the difference in delay across overfitting prevention approaches, we performed the Mann-Whitney U test~\citep{Mann1947OnAT} at a significance level of $\alpha=0.05$ to determine whether the distributions of the delay epochs of early stopping and our approach are significantly different. We also computed Cliff's delta~$d$~\citep{Cliff} effect size to quantify the difference based on the provided thresholds~\citep{Cliff_threshold}.

% \begin{equation} \label{effectsize}
% \mathrm{Effect \ size} = 
% \left\{
% \begin{array}{ll}
% 	negligible,  & \mathrm{if} \ |d|  \le 0.147 \\
% 	small,  & \mathrm{if} \ 0.147 < |d|  \le 0.33 \\
% 	medium,  & \mathrm{if} \ 0.33 < |d|  \le 0.474 \\
% 	large,  & \mathrm{if} \ 0.474 < |d|  \le 1 \\
% \end{array}\right.
% \end{equation}

\begin{figure}[t]
    \centering
    \includegraphics[width=\columnwidth]{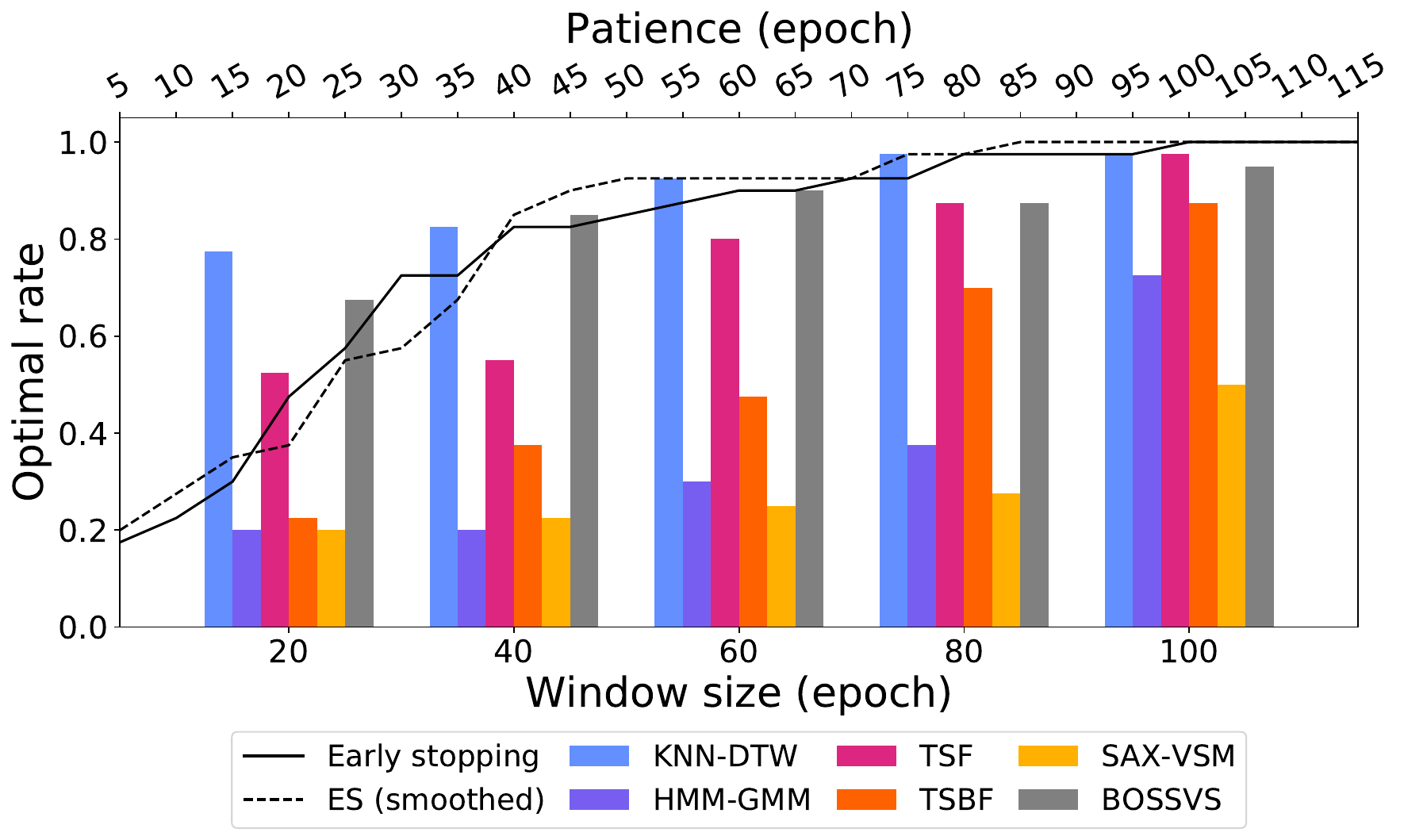}
    \caption{The optimal rate of our overfitting prevention approach~(using a rolling window) and early stopping with different patience values.}
    \label{fig:cmp_correct_rate}
\end{figure}

\begin{table*}[t]
\caption{The median delay and average accuracy of early stopping (es) and our overfitting prevention approaches (using a rolling window) with different window sizes (ws).}
\label{tab:stop_delay_epoch}
\begin{center}
\begin{tabular}{lrrrrrlrrrrr}
\toprule
\multirow{2}{*}{\begin{tabular}[c]{@{}l@{}}Prevention\\ approach\end{tabular}} &
  \multicolumn{5}{c}{Median delay / ws} &
   &
  \multicolumn{5}{c}{Average accuracy / ws} \\ \cmidrule(lr){2-6} \cmidrule(l){8-12} 
 &
  20 &
  40 &
  60 &
  80 &
  100 &
   &
  20 &
  40 &
  60 &
  80 &
  100 \\ \midrule
ES &
  \multicolumn{1}{l}{20.0} &
  40.0 &
  60.0 &
  80.0 &
  100.0 &
   &
  0.40 &
  0.42 &
  0.42 &
  0.43 &
  0.43 \\
ES (smoothed) &
  \multicolumn{1}{l}{20.0} &
  43.5 &
  62.0 &
  82.0 &
  97.5 &
   &
  0.39 &
  0.42 &
  0.43 &
  0.43 &
  0.43 \\ \midrule
KNN-DTW &
  \textbf{31.0} &
  \textbf{27.0} &
  \textbf{37.5} &
  \textbf{42.5} &
  \textbf{45.5} &
   &
  \textbf{0.42} &
  \textbf{0.42} &
  \textbf{0.43} &
  \textbf{0.43} &
  \textbf{0.43} \\
HMM-GMM &
  5.0 &
  6.5 &
  16.5 &
  28.0 &
  41.5 &
   &
  0.37 &
  0.37 &
  0.38 &
  0.39 &
  0.41 \\
TSF &
  12.5 &
  22.0 &
  31.0 &
  39.5 &
  44.0 &
   &
  0.40 &
  0.40 &
  0.42 &
  0.43 &
  0.43 \\
TSBF &
  8.5 &
  15.0 &
  25.0 &
  37.0 &
  47.0 &
   &
  0.38 &
  0.39 &
  0.40 &
  0.41 &
  0.42 \\
BOSSVS &
  7.0 &
  29.0 &
  34.5 &
  48.5 &
  56.5 &
   &
  \multicolumn{1}{l}{0.42} &
  \multicolumn{1}{l}{0.42} &
  \multicolumn{1}{l}{0.43} &
  \multicolumn{1}{l}{0.43} &
  \multicolumn{1}{l}{0.43} \\
SAX-VSM &
  4.0 &
  11.0 &
  9.5 &
  16.0 &
  24.0 &
   &
  \multicolumn{1}{l}{0.37} &
  \multicolumn{1}{l}{0.37} &
  \multicolumn{1}{l}{0.38} &
  \multicolumn{1}{l}{0.38} &
  \multicolumn{1}{l}{0.40} \\ \bottomrule
\end{tabular}
\end{center}
\end{table*}

\results
\textit{Our proposed approach, utilizing KNN-DTW with both rolling window and whole observed history, has a similar or higher optimal rate than early stopping for overfitting prevention}. Other studied classifiers do not perform as well as KNN-DTW for overfitting prevention. As Figure~\ref{fig:cmp_correct_rate} and Table~\ref{tab:whole_his_prevention} show, using KNN-DTW with either a rolling window or the whole observed history outperforms early stopping at identifying the optimal epoch. In particular, our approach with KNN-DTW based on the rolling window has a higher or the same optimal rate as both early stopping approaches when using up to 80 epochs as the patience parameter and window size. For example, our approach with KNN-DTW obtains 78\% optimal rate when setting the window size to 20 epochs, while both early stopping approaches achieve less than 50\% optimal rate when the patience parameter is set to the same epochs. However, when the patience parameter is greater than 80 epochs, both early stopping approaches can identify almost all of the optimal epochs. The reason is that 90\% of the training histories in the real-world dataset have around 200 epochs, hence, a large patience value makes it easy for early stopping to choose the optimal epoch. In addition, Table~\ref{tab:whole_his_prevention} shows that our approach with KNN-DTW based on the whole observed history also obtains a higher optimal rate compared to both early stopping approaches. For example, the KNN-DTW classifier achieves a 95\% optimal rate with a median delay of 43.5 epochs, whereas early stopping approaches achieve around 85\% using the same number~(i.e., between 40 to 45 epochs patience in Figure~\ref{fig:cmp_correct_rate}). %Furthermore, our method using KNN-DTW can be used with classification error (i.e., zero-one loss) and still outperforms early stopping~(see details in Appendix~\ref{appendix:acc_curve}). 

\begin{figure}[t]
     \centering
     \includegraphics[width=\columnwidth]{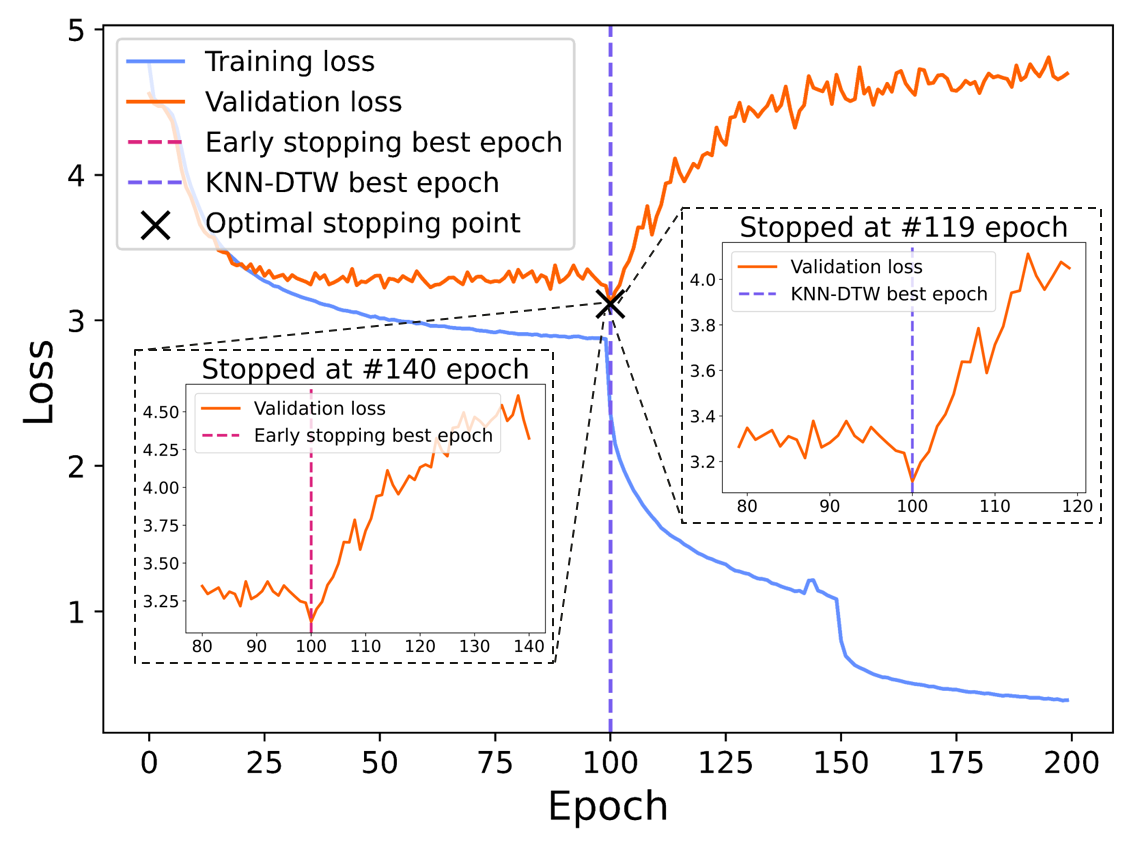}
     \caption{Our approach using KNN-DTW (set the window size as 40 epochs) stops earlier than the early stopping (set the patience parameter as 40 epochs) but both achieve the same optimal epoch.}
     \label{fig:cmp_stop_earlier}
\end{figure}

% \input{tabs/sig_test_loss}

% \begin{figure*}[htb]
% \begin{center}
% \includegraphics[width=0.8\textwidth]{es_smooth_example_bad}
% \end{center}
% \caption{An example in which early stopping cannot achieve the optimal epoch based on smoothed validation loss (the patience parameter is set to 20 epochs and the smoothing window size is 5).}
% \label{fig:es_smooth_example_bad}
% \end{figure*}

\textit{Our approach using KNN-DTW and a rolling window can stop training a DL model earlier than early stopping (based on both smoothed and non-smoothed validation loss) with the same or higher accuracy.} As shown in Table~\ref{tab:stop_delay_epoch}, with the same number of epochs for the patience parameter and window size, our approach can save training time (i.e., reducing delay between the stopped epoch and the best epoch) compared to early stopping, except for a window size of 20 epochs. For instance, when setting both the patience parameter and window size to 40 epochs, KNN-DTW and early stopping have the same average accuracy, but KNN-DTW has a median delay of 27 epochs while early stopping has a fixed delay of 40 epochs. The significance test results %in Table~\ref{tab:sig_test_loss} 
indicate that the delay difference between KNN-DTW and early stopping is significant (except when using a window size of 20), with a medium to large effect size. Furthermore, early stopping with smoothed loss curves has a median delay of 43.5, which is slower than using the original loss curves with the same patience parameter (40 epochs). Early stopping using the smoothed loss could hurt the performance of early stopping%, as shown in Figure~\ref{fig:es_smooth_example_bad}, where early stopping misses the optimal epoch due to the smoothed curve. Smoothing does not improve early stopping's performance 
and cannot compete with our approach. In comparison to the delay in early stopping, the delay between the stopped epoch and the best epoch is at least 32\% shorter with our approach using KNN-DTW. Figure~\ref{fig:cmp_stop_earlier} provides an example where both early stopping and our approach identify the optimal epoch, but our approach stops 21 epochs earlier than early stopping~(which stops with a 40 epochs delay).

\textit{Among our two approaches for overfitting prevention, we recommend using KNN-DTW with a rolling window.} Although using the whole observed history may achieve a higher optimal rate than using a rolling window for our approach, we note that we can predict the optimal epoch much earlier with the rolling window approach for a very small trade-off in optimal rate (with a similar average accuracy). As shown in Table~\ref{tab:stop_delay_epoch} and Figure~\ref{fig:cmp_correct_rate}, our approach with KNN-DTW achieves an 83\% optimal rate with a median delay of 27 epochs and a 90\% optimal rate with a median delay of 37.5 epochs using the window size as 40 and 60 epochs respectively. However, the median delay of KNN-DTW when using the whole observed history is 43.5, while the rolling window approach with a window size of 80 or more epochs can achieve a higher optimal rate~(98\% vs. 95\% accuracy) with a shorter delay~(42.5 vs. 43.5 epochs). In summary, we suggest using the rolling window approach since it stops earlier with a relatively small optimal rate drop using a small window~(e.g., 40 epochs) and outperforms the whole observed history approach when using a large window size~(e.g., 80 epochs). 

% \begin{figure}[htb]
% \begin{center}
% \includegraphics[width=0.9\columnwidth]{es_smooth_example_good}
% \end{center}
% \caption{An example in which early stopping achieves the optimal epoch based on smoothed validation loss (the patience parameter is set to 40 epochs and the smoothing window size is 5).}
% \label{fig:es_smooth_example_good}
% \end{figure}
% Figure~\ref{fig:es_smooth_example_good} shows a case where smoothing helps early stopping achieve the optimal epoch, but this is not a common scenario.

% \smallskip
% \begin{Summary}{\ Summary of RQ2}{}

\hypobox{
	\emp{RQ2 Takeaway:} 
Our proposed approach using KNN-DTW with a rolling window or whole observed history outperforms early stopping for overfitting prevention and can stop training DL models earlier with the same or higher accuracy. Among our two approaches, we recommend using KNN-DTW with a rolling window for early stopping, which achieves a high optimal rate with a shorter delay.
}
% \end{Summary}
% \input{sections/guidelines}
\section{Threats to validity}\label{sec:threats}

% \smallskip\noindent\textbf{Construct Validity} 
\subsection{Construct Validity}
The construct validity of our approach may be affected by the manual labelling process for the simulated training dataset used in overfitting detection. The definition of overfitting is an abstract concept and may result in ambiguity or disagreements among authors. To mitigate this threat, two authors labelled the training histories independently, achieving a 95\% agreement rate. Following this, the authors engaged in multiple rounds of discussions to resolve any disagreements (as detailed in Section~\ref{sec:simulated_dataset}). Despite these efforts, some subjectivity in the labelling process might impact the validity of our results.

Another potential threat to construct validity is the choice of the monitoring metric used in overfitting prevention. Although validation loss is a widely used metric for monitoring the DL model performance during the training process, different DL tasks may require alternative metrics. We conducted additional experiments using classification error (i.e., zero-one loss) for overfitting prevention, and our approach using KNN-DTW still outperformed early stopping.
% We conducted additional experiments in Appendix~\ref{appendix:acc_curve} using classification error (i.e., zero-one loss) for overfitting prevention, and our approach using KNN-DTW still outperformed early stopping.

% \smallskip\noindent\textbf{Internal Validity}
\subsection{Internal Validity}
Our proposed approach relies on the assumption that overfitting can be detected and prevented through the analysis of DL model training histories. However, certain cases of overfitting may not be captured by examining the training histories alone. For instance, data leakage caused by data augmentation or preprocessing in the entire dataset before data splitting~(into training, validation, and test sets) could lead to overfitting, but detecting or preventing it solely by inspecting the training history would be challenging.

% \smallskip\noindent\textbf{External validity}
\subsection{External Validity}
We evaluated our proposed approach using a real-world dataset that contains training histories from top AI venues. However, it is still possible that the approach may not generalize well to all types of DL models or datasets. Secondly, our real-world evaluation is based on only 40 data points (of which 29 training histories belong to overfit models), which might not be enough data points to claim generalizability of our proposed approach. Please note that collecting authoritative examples of overfit training history is very hard since researchers and practitioners typically do not report the training history of models that were overfit. In addition, collecting these data points requires one to replicate the studies that report overfit models, which is a very time and resource intensive task. Hence we were limited to 40 data points for the real-world dataset in our study. However, we invite future research to verify the validity of \OverfitGuard using our replication package on their own DL model training histories. In addition, the computational resources required to use the proposed approach for inference could limit its applicability in specific situations. For instance, the increased computational cost may be prohibitive in environments with constrained computational resources, while our approach demonstrates improved performance in overfitting detection and prevention than existing approaches.

\section{Conclusion and future work}\label{sec:conclusion}

In this paper, we propose a non-intrusive overfitting detection and prevention approach using time series classifiers trained on the training history of DL models. Our approach (when using the KNN-DTW time series classifier) has (1)~better classification performance than correlation-based approaches for overfitting detection, and (2)~greater accuracy than early stopping for overfitting prevention with a shorter delay. We evaluate our approach on a real-world dataset of labelled training histories collected from the papers published at top AI venues in the last 5 years. Our approach can be a useful tool for researchers and developers of DL software. We have shared the trained time series classifiers in the replication package for reuse, along with all of the training histories and labels. One limitation of our approach is that our best-performing time series classifier takes longer to perform the inference required to detect and prevent overfit than the studied baselines. We encourage future work to optimize time series classifiers to enable overfitting detection and prevention in real-time with smaller delays. We also encourage the future work to investigate if adding more real world examples to \OverfitGuard's training data or using online training to constantly update \OverfitGuard would improve its detection and prevention performance.

\bibliographystyle{IEEEtranSN}
\bibliography{mybib}

\end{document}